\documentclass[12pt]{article}
\usepackage{amssymb,amsmath,epsfig}

\begin{document}

\title{\bf Stability Analysis of Thin-Shell Wormholes from Charged Black String}

\author{M. Sharif$^1$ \thanks{msharif.math@pu.edu.pk} and M. Azam$^{1,2}$
\thanks{azammath@gmail.com}\\
$^1$ Department of Mathematics, University of the Punjab,\\
Quaid-e-Azam Campus, Lahore-54590, Pakistan.\\
$^2$ Division of Science and Technology, University of Education,\\
Township Campus, Lahore-54590, Pakistan.}

\date{}

\maketitle
\begin{abstract}
In this paper, we construct thin-shell wormholes from charged black
string through cut and paste procedure and investigate its
stability. We assume modified generalized Chaplygin gas as a dark
energy fluid (exotic matter) present in the thin layer of
matter-shell. The stability of these constructed thin-shell
wormholes is investigated in the scenario of linear perturbations.
We conclude that static stable as well as unstable configurations
are possible for cylindrical thin-shell wormholes.
\end{abstract}
{\bf Keywords:} Israel junction conditions; Stability; Black strings.\\
{\bf PACS:} 04.20.Gz; 04.40.Nr; 04.70.Bw.

\section{Introduction}

Wormholes are the hypothetical objects having a peculiar property of
containing exotic matter (which violates null energy condition). The
first ever wormhole model was known to be the Einstein-Rosen bridge
\cite{1}, which was obtained as a part of the maximally extended
Schwarzschild solution. The main problem with this wormhole is the
existence of event horizon which prevents observers to move freely
from one universe to the other. Later, Morris and Thorne \cite{2}
presented the first traversable Lorentzian wormhole as solution of
the Einstein field equations. The key feature of this wormhole is
that it does not contain event horizon and an observer may freely
move in both universes through a handel (tunnel) known as wormhole
throat.

Traversable wormholes have some issues such as their mechanical
stability, the unavoidable amount of exotic matter present at the
wormhole throat, etc. The violation of energy conditions due to the
presence of exotic matter in these configurations is a debatable
issue in general relativity which is the main hurdle in its
observational evidence. To minimize the violation of energy
conditions, Visser \cite{3,4} used the cut and paste technique on a
black hole to build a thin-shell wormhole. He used the
Darmois-Israel formalism \cite{5,6} to study the dynamical behavior
of thin-shell wormhole made of two identical geometries.

Many authors studied the stability of thin-shell wormholes against
linear perturbations through a standard potential approach. Poisson
and Visser \cite{6a} explored the stability of the Schwarzschild
thin-shell wormhole. Eiroa and Romero \cite{7} generalized this
analysis for the Reissner-Nordstr\"{o}m thin-shell wormholes, while
Lobo and Crawford \cite{8} included the cosmological constant for
the same analysis. For the sake of stable thin-shell wormhole
configurations, people have also studied wormhole solutions in
modified theories of gravity. For instance, Thibeault et al.
\cite{9} found stable thin-shell wormhole in Einstein-Maxwell theory
with a Gauss-Bonnet term. Rahaman et al. \cite{10,10a} explored
thin-shell wormhole solutions in heterotic string theory and in the
Randall-Sundrum scenario. Mazharimousavi et al. \cite{11,11a} found
viable thin-shell wormhole solutions in the Einstein-Hoffmann
Born-Infeld theory and Einstein-Yang-Mills Dilaton gravity. In
recent papers \cite{11b}, we have investigated the stability of
cylindrical and spherical geometries in Newtonian and post-Newtonian
approximations and also spherically symmetric thin-shell wormholes.

Thorne \cite {12} emphasized that models with cylindrical symmetry
are the ideal one. These objects have widely been used to study
cosmic strings \cite{12a} which play a vital role in different
physical phenomena like gravitational lensing, galaxy formation,
thin-shell wormholes, etc. Some literature \cite{13}-\cite{16}
indicates keen interest in the study of cylindrical thin-shell
wormholes. In a sequence of papers \cite{17}-\cite{20}, the
stability of cylindrical thin-shell wormholes associated with local
and global cosmic strings have been studied. It was found that the
wormhole throat would expand or collapse according to the velocity
sign and stable cylindrical thin-shell wormhole configurations could
not be possible. Recently, we have explored stable thin-shell
wormhole configurations supported by Chaplygin equation of state
\cite{a}.

In this paper, we construct thin-shell wormholes from charged black
string using cut and paste procedure and investigate its stability
through linear perturbations. We consider the Darmois-Israel
formalism for the dynamical analysis of the system with modified
generalized Chaplygin gas (MGCG) to matter shell. The paper is
organized as follows. Section \textbf{2} deals with the general
formalism for the construction of thin-shell wormholes. In section
\textbf{3}, we discuss the linearized stability analysis and apply
to charged black string thin-shell wormholes. In the last section,
we summarize our results.

\section{Thin-Shell Wormholes: General Formalism}

The charged static cylindrically symmetric spacetime is given by
\cite{21}
\begin{equation}\label{3}
ds^2=-\Phi(r)dt^{2}+\Phi^{-1}(r)dr^{2}+h(r)(d\phi^{2}+\alpha^2{dz^2}),
\end{equation}
where
$$\Phi(r)=\left(\alpha^2r^2-\frac{4M}{\alpha{r}}+\frac{4Q^2}{\alpha^2r^2}\right),~h(r)=r^2,$$
with the following constraints on the coordinates
$$-\infty<t<\infty,\quad 0< {r}<\infty, \quad
-\infty<{z}<{\infty},\quad 0\leq{\phi}\leq{2\pi}.$$ Here, the
parameters $M,~Q$ are the ADM mass and charge density, respectively
and $\alpha=-\frac{\Lambda}{3}>0$, $\Lambda$ is the cosmological
constant. The inner and outer event horizons of the charged black
string are given as
\begin{equation}\label{4}
r_{\pm}=\frac{(4M)^\frac{1}{3}}{2\alpha}\left[\sqrt{s}\pm
\sqrt{{2}\sqrt{s^2-Q^2\left(\frac{2}{M}\right)^\frac{4}{3}}-s}\right],
\end{equation}
provided that the inequality $Q^2\leq\frac{3}{4}M^\frac{4}{3}$
holds, where $s$ is given by
\begin{eqnarray}\label{5}
s&=&\left(\frac{1}{2}+\frac{1}{2}\sqrt{1-\frac{64Q^6}{27M^4}}
\right)^\frac{1}{3}
+\left(\frac{1}{2}-\frac{1}{2}\sqrt{1-\frac{64Q^6}{27M^4}}
\right)^\frac{1}{3}.
\end{eqnarray}
For $Q^2>\frac{3}{4}M^\frac{4}{3}$, the given metric has no event
horizon and represents a naked singularity. If
$Q^2=\frac{3}{4}M^\frac{4}{3}$, the inner and outer horizons merge,
which corresponds to the extremal black strings.

We follow the Darmois-Israel formulation \cite{5,6} for the
dynamical analysis of mathematically constructed thin-shell
wormholes. For this purpose, we assume radius $``a"$ greater than
event horizon $r_h$ to avoid singularities and horizons in wormhole
configuration. We take two identical copies $\mathcal{W}^{\pm}$ with
$r\geq{a}$ of the cylindrical vacuum solution defined as
\begin{equation}\label{6}
\mathcal{W}^{\pm}=\{x^{\mu}=(t,r,\phi,z)/r\geq{a}\}.
\end{equation}
We join these geometries at the timelike hypersurface
$\Sigma=\Sigma^\pm=\{r-a=0\}$ to get a geodesically complete
manifold, i.e., $\mathcal{W}=\mathcal{W}^{+}\cup{\mathcal{W}^{-}}$
satisfying the radial flare-out condition, i.e., $h'(a)=2a>0$
\cite{16}. The two regions are connected at the surface $\Sigma$
(surface of minimal area) with the throat radius $a$. The induced
metric at the throat $\Sigma$ with coordinates $\eta^i=(\tau, \phi,
z)$ is defined as
\begin{equation}\label{7}
ds^2=-d\tau^2+a^2(\tau)(d\phi^2+\alpha^2dz^2).
\end{equation}
We take throat radius $a$ as a function of $\tau$ to understand the
dynamics of the thin-shell wormhole. The presence of thin layer of
matter at the shell leads to discontinuity in the extrinsic
curvatures across a junction surface, where
$K^{+}_{ij}-K^{-}_{ij}=\kappa_{ij}$, and the extrinsic curvature
$K^{\pm}_{ij}$ is defined on $\Sigma$
\begin{equation}\label{8}
K^{\pm}_{ij}=-n^{\pm}_{\gamma}\left(\frac{{\partial}^2x^{\gamma}_{\pm}}
{{\partial}{\eta}^i{\partial}{\eta}^j}+{\Gamma}^{\gamma}_{{\mu}{\nu}}
\frac{{{\partial}x^{\mu}_{\pm}}{{\partial}x^{\nu}_{\pm}}}
{{\partial}{\eta}^i{\partial}{\eta}^j}\right),\quad(i, j=0,2,3).
\end{equation}
The $4$-vector unit normals $n^{\pm}_{\gamma}$ to
$\mathcal{W}^{\pm}$ are
\begin{equation}\label{9}
n^{\pm}_{\gamma}=\pm\left|g^{\mu\nu}\frac{\partial{f}}{\partial{x^{\mu}}}
\frac{\partial{f}}{\partial{x^{\nu}}}\right|
=\left(-\dot{a},\frac{\sqrt{\Phi(r)+\dot{a}^2}}{\Phi(r)},0,0\right),
\end{equation}
satisfying the relation $n^{\gamma}n_{\gamma}=1$. Using
Eqs.(\ref{3}) and (\ref{8}), the non-trivial components of the
extrinsic curvature are
\begin{equation}\label{10}
K^{\pm}_{\tau\tau}=\mp\frac{\Phi'(a)+2\ddot{a}}{2\sqrt{\Phi(a)+\dot{a}^2}},
\quad K^{\pm}_{\phi\phi}= \pm
\frac{1}{a}\sqrt{\Phi(a)+\dot{a}^2},\quad
K^{\pm}_{zz}=\alpha^2K^{\pm}_{\phi\phi},
\end{equation}
where dot and prime mean derivative with respect to $\tau$ and $r$,
respectively.

Now using the relations between the extrinsic curvatures
$$[K_{ij}]=K^{+}_{ij}-K^{-}_{ij},\quad K=tr[K_{ij}]=[K^{i}_{i}],$$
the Einstein equations (called Lanczos equations) are defined on the
shell as
\begin{equation}\label{11}
{S_{ij}}=\frac{1}{8\pi}\left\{g_{ij}K-[K_{ij}]\right\},
\end{equation}
where $S_{ij}=diag(\sigma,p_\phi,p_z)$ is the surface
energy-momentum tensor, $\sigma$ and $p_\phi,~p_z$ are the surface
energy density and surface pressures, respectively. With
Eqs.(\ref{10}) and (\ref{11}), we obtain
\begin{eqnarray}\label{12}
\sigma&=&-\frac{1}{2\pi{a}}\sqrt{\Phi(a)+\dot{a}^2},\\\label{12a}
p&=&p_{\phi}=p_z=\frac{1}{8\pi{a}}\frac{2a\ddot{a}+2\dot{a}^2
+2\Phi(a)+a\Phi'(a)}{\sqrt{\Phi(a)+\dot{a}^2}}.
\end{eqnarray}
The negative surface energy density (\ref{12}) supports the presence
of exotic matter at the throat.

For the dynamical characterization of the shell, we consider the
MGCG as exotic matter on the shell. The equation of state for MGCG
is defined as
\begin{equation}\label{13}
p=A{\sigma}-\frac{B}{\sigma^\beta},
\end{equation}
where $A,~B$ are positive constants and $0<{\beta}\leq1$. This
equation combines various equations of state and reduces to the
following classes for different values of the parameters $A,~B$ and
$\beta$, such as
\begin{itemize}
\item{for $A=0,~\beta=1$, it corresponds to the usual Chaplygin gas.}
\item{for $A=0$, it corresponds to pure generalized Chaplygin gas (GCG).}
\item{for $\beta=1$, it is another form of modified Chaplygin gas (MCG).}
\end{itemize}
Ujjal \cite{u} have generalized MGCG to variable MGCG by assuming
$B$ as a function of the scale factor $a$, i.e.,
$B=B(a)=B_{0}a^{-m}$, where $B_0,~m$ are the positive constants. In
this work, we have assumed $B$ as a positive constant. Inserting
Eqs.(\ref{12}) and (\ref{12a}) in (\ref{13}), we obtain a second
order differential equation describing the evolution of the wormhole
throat
\begin{eqnarray}\label{14}
&&\left\{\left[2\ddot{a}+\Phi'(a)\right]a^2+\left[\left(\Phi(a)+\dot{a}^2\right)
\left(1+2A\right)\right]2a\right\}\left[2a\right]^{\beta}\\\nonumber&-&2B(4\pi{a^2})^{1+\beta}
\left[\Phi(a)+\dot{a}^2\right]^\frac{1-\beta}{2}=0.
\end{eqnarray}

\section{Linearized Stability Analysis: A Standard Approach}

In this section, we analyze the stability of static solutions of
thin-shell wormhole under the standard potential approach
\cite{6a,7}. For this purpose, the static configuration of surface
energy density, surface pressure and dynamical equation for the
thin-shell wormhole yields
\begin{eqnarray}\label{15}
\sigma_0=-\frac{\sqrt{\Phi(a_0)}}{2\pi{a_0}},\quad
p_0=\frac{2\Phi(a_0)+a_0\Phi'(a_0)}{8\pi{a_0}\sqrt{\Phi(a_0)}},
\end{eqnarray}

\begin{eqnarray}\label{16}
\left\{a^2_0\Phi'(a_0)+2a_0
\left(1+2A\right)\Phi(a_0)\right\}\left[2a_0\right]^{\beta}-2B(4\pi{a^2_0})^{1+\beta}
\left[\Phi(a_0)\right]^\frac{1-\beta}{2}=0.
\end{eqnarray}
The surface energy density and pressure satisfy the conservation
equation
\begin{eqnarray}\label{17}
\frac{d}{d\tau}(\sigma{\Omega})+p\frac{d\Omega}{d\tau}=0,
\end{eqnarray}
where $\Omega=4\pi{a^2}$ is the area of the wormhole throat. This
equation describes the change in internal energy of the throat plus
the work done by the throat's internal forces. We can write this
equation as follows
\begin{eqnarray}\label{18}
\dot{\sigma}=-2(\sigma+p)\frac{\dot{a}}{a}.
\end{eqnarray}
Defining ${\sigma}'=\frac{\dot{\sigma}}{\dot{a}}$, this equation
takes the form
\begin{equation}\label{19}
a{\sigma}'=-2(\sigma+p).
\end{equation}

For the stability of static configuration under the radial
perturbations around $a=a_0$, we rearrange Eq.(\ref{12}) to obtain
the thin-shell equation of motion
\begin{equation}\label{20}
\dot{a}^2+V(a)=0.
\end{equation}
This completely determines the dynamics of the thin-shell wormhole,
where $V(a)$ is known as potential function given by
\begin{equation}\label{21}
V(a)=\Phi(a)-\left[2\pi{a}{\sigma(a)}\right]^2.
\end{equation}
The stability of static solutions requires $V''(a_0)>0$,
$V(a_0)=0=V'(a_0)$. For this purpose, we apply the Taylor series
expansion to $V(a)$ upto second order around $a_0$
\begin{eqnarray}\label{22}
V(a)=V(a_0)+V'(a_0)(a-a_0)+\frac{1}{2}V''(a_0)(a-a_0)^2+O[(a-a_0)^3].
\end{eqnarray}
Taking the first derivative of Eq.(\ref{21}) and using (\ref{19}),
we obtain
\begin{equation}\label{23}
V'(a)=\Phi'(a)+8{\pi}^2a\sigma(a)\left[\sigma(a)+p(a)\right].
\end{equation}
We have another useful relation from the equation of state
\begin{equation}\label{24}
p'(a)={\sigma'(a)}\left[(1+\beta)A-\frac{\beta{p(a)}}{\sigma(a)}\right],
\end{equation}
which may then be written as
\begin{equation}\label{25}
{\sigma'(a)}+2p'(a)={\sigma}'(a)\left[1+2\{(1+\beta)A-\frac{\beta{p(a)}}{\sigma(a)}\}\right].
\end{equation}
The second derivative of potential function along with the above
equation leads to
\begin{eqnarray}\nonumber
V''(a)&=&\Phi''(a)-8{\pi}^2\left\{[\sigma(a)+2p(a)]^2+2\sigma(a)
\left[\sigma(a)+p(a)\right]\right.\\\label{26}&\times&\left.\left[1+2\left((1+\beta)A
-\frac{\beta{p(a)}}{\sigma(a)}\right)\right]\right\}.
\end{eqnarray}
Using Eq.(\ref{15}) both $V(a)$ and $V'(a)$ vanish at $a=a_0$, while
$V''(a_0)$ becomes
\begin{eqnarray}\nonumber
V''(a_0)&=&\Phi''(a_0)+\frac{(\beta-1){\Phi'(a_0)}^2}{2\Phi(a_0)}+\frac{{\Phi'(a_0)}}{a_0}
\left[1+2(1+\beta)A\right]\\\label{27}&-&\frac{2{\Phi(a_0)}(1+\beta)}{a_0}\left(1+A\right).
\end{eqnarray}

\subsection{Charged Black String Thin-Shell Wormholes}

In this section, we formulate the charged black string thin-shell
wormholes and discuss the stability of their static solutions. The
surface energy density and pressure for the charged black string
wormhole with Eq.(\ref{15}) becomes
\begin{eqnarray}\label{28}
\sigma_0=-\frac{\sqrt{\alpha^4{a^4_0}-4M\alpha{a_0}+4Q^2}}{2\pi\alpha{a_0}},\quad
p_0=\frac{\alpha^3{a^3_0}-4M}{2\pi{a_0}\sqrt{\alpha^4{a^4_0}-4M\alpha{a_0}+4Q^2}}.
\end{eqnarray}
Using these values in Eqs.(\ref{13}) and (\ref{26}), the dynamical
equation and the second derivative of potential for the thin-shell
wormhole satisfied by the throat radius becomes
\begin{eqnarray}\nonumber
&&\alpha^4{a^4_0}-M\alpha{a_0}+A\left(\alpha^4{a^4_0}-4M\alpha{a_0}+4Q^2\right)
-B\left(2\pi\alpha{a^2_0}\right)^{1+\beta}
\\\label{29}&\times&\left(\alpha^4{a^4_0}-4M\alpha{a_0}+4Q^2\right)^{\frac{1-\beta}{2}}=0,
\end{eqnarray}
and
\begin{eqnarray}\nonumber
V''(a_0)&=&\frac{4}{\alpha^2a^4_0(\alpha^4a^4_0-4\alpha{a_0}M+4Q^2)}
\left\{(1+\beta)\left[-6\alpha^2a^2_0{M^2}\right.\right.\\\nonumber
&\times&\left.\left.(1+4A)-32AQ^2+(1+A)\alpha^5a^5_0M+8\alpha{a_0}MQ^2(1+7A)
\right.\right.\\\label{30}
&-&\left.\left.8(1+A)\alpha^4{a^4_0}Q^2\right]+16\alpha^4{a^4_0}Q^2-9\alpha^5{a^5_0}M\right\}.
\end{eqnarray}
\begin{figure}
\centering \epsfig{file=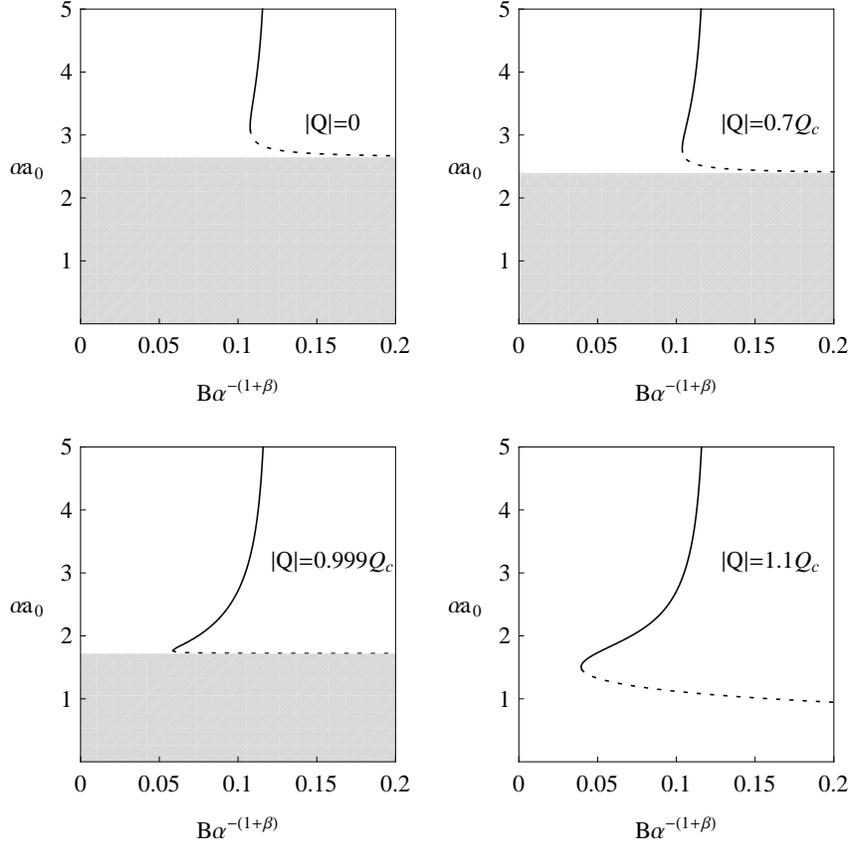}\caption{Charged black string
thin-shell wormholes for $\alpha=0.6,~A=M=1$ and $\beta=0.2$ for
different values of charge. The solid and dotted curves correspond
to the stable and unstable static solutions under linear
perturbations. The non-physical regions are represented by the
shaded regions (grey zones), where $a_0\leq{r_h}$.}
\end{figure}
\begin{figure}
\centering \epsfig{file=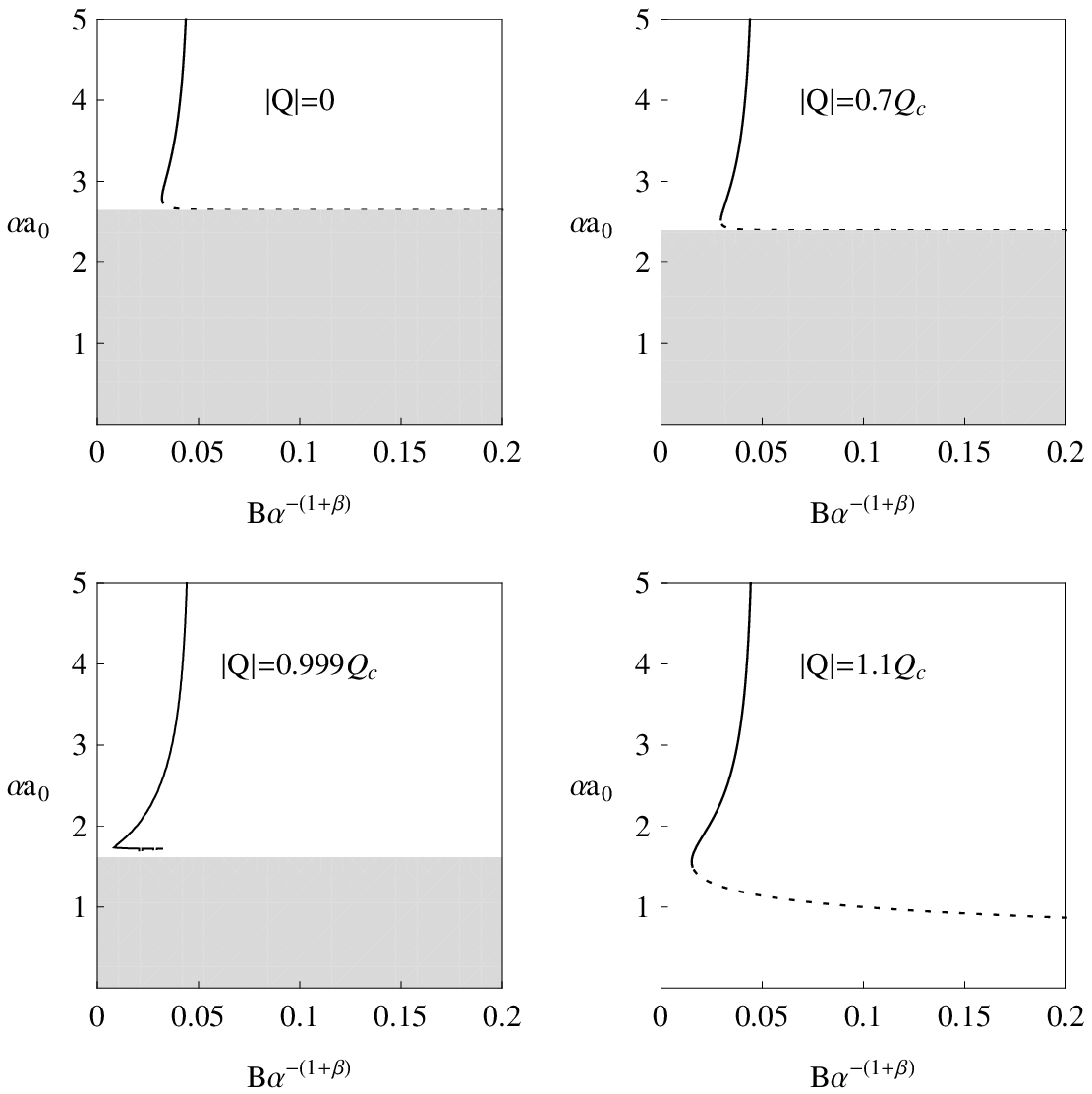}\caption{Charged black string
thin-shell wormholes for $\alpha=0.6,~A=M=1$ and $\beta=0.6$.}
\end{figure}

Now we explore the nature of static solutions whether they are
stable or unstable. The existence of static solution is constrained
by the condition $a_0>{r_h}$, i.e., throat radius must be greater
than the horizon radius. On the other hand, for $a_0\leq{r_h}$, the
static solutions do not exist which correspond to the non-physical
zone (grey zone) as shown in Figures \textbf{1-6}. Due to the
complicated nature of Eq.(\ref{29}), we solve this equation
numerically and find $a_0$ for $\beta=0.2,~0.6,~1$ and then replace
the solution in Eq.(\ref{30}). If $V''(a_0)>0$, we have stable
static solution which is represented by the solid curve, whereas for
$V''(a_0)<0$, the solution is unstable represented by the dotted
curve. The behavior of static solutions depends upon the critical
value of charge, $Q_c=0.866025$. We can discuss the solutions given
in Figures \textbf{1-2} for $\beta=0.2,~0.6$ as follows:
\begin{itemize}
\item{For $|Q|=0$, there exist both stable and unstable solutions for the black string
thin-shell wormhole. The unstable solution approaches to the horizon
radius for large values of $B\alpha^{-(1+\beta)}.$}
\item{For $|Q|=0.7Q_c$, this gives similar behavior as for the case $|Q|=0$.}
\item{For $|Q|=0.999Q_c$, i.e., $|Q|$ is nearly equal to the value of the critical charge.
When $\beta=0.2$, both stable and unstable solutions exist, while
for $\beta=0.6$, stable solution exists only for small values of
$B\alpha^{-(1+\beta)}$. Moreover, the horizon radius decreases with
the increase of charge.}
\item{When $|Q|=1.1Q_c$, i.e., $|Q|$ has greater value than the critical value of the charge.
In this case, stable and unstable solutions exist in each case for
the increasing value of $B\alpha^{-(1+\beta)}$ and the horizon
radius gradually disappears for $|Q|>Q_c$.}
\end{itemize}
\begin{figure}
\centering \epsfig{file=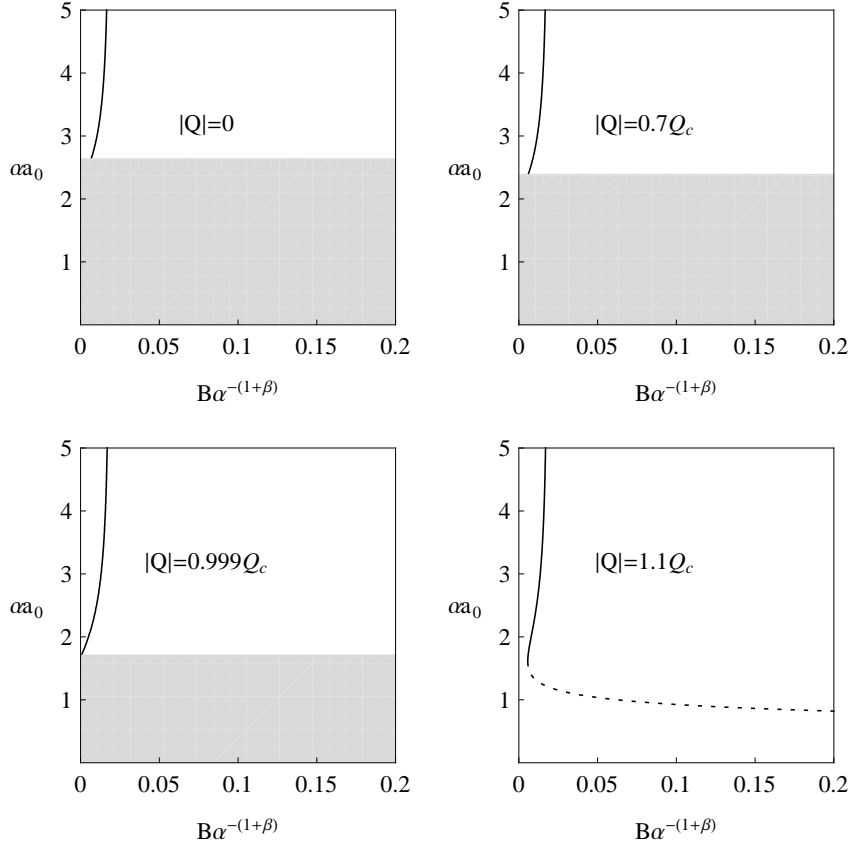}\caption{Charged black string
thin-shell wormholes correspond to the parameters
$\alpha=0.6,~A=M=1$ and $\beta=1$.}
\end{figure}

We also explore the stability of static solutions corresponding to
$\beta=1,$ which corresponds to the MCG as shown in Figure
\textbf{3}. Notice that for $|Q|<Q_c,$ there always exists a stable
static solution for small values of $B\alpha^{-(1+\beta)}$ and
vanishes for large values of $B\alpha^{-(1+\beta)}$. When $|Q|>Q_c$,
we again have two solutions stable for small values of
$B\alpha^{-(1+\beta)}$ and unstable for large values of
$B\alpha^{-(1+\beta)}$. Also, similar to the above cases, the
horizon radius decreases and eventually disappears for the
increasing value of $|Q|$.

Now we analyze the stability of static solutions which correspond to
GCG and usual Chaplygin gas. For this purpose, we take $A=0$ and
$\beta=0.2,~0.6,~1$ in Eq.(\ref{13}) and results are shown in
Figures \textbf{4-6}.
\begin{figure}
\centering \epsfig{file=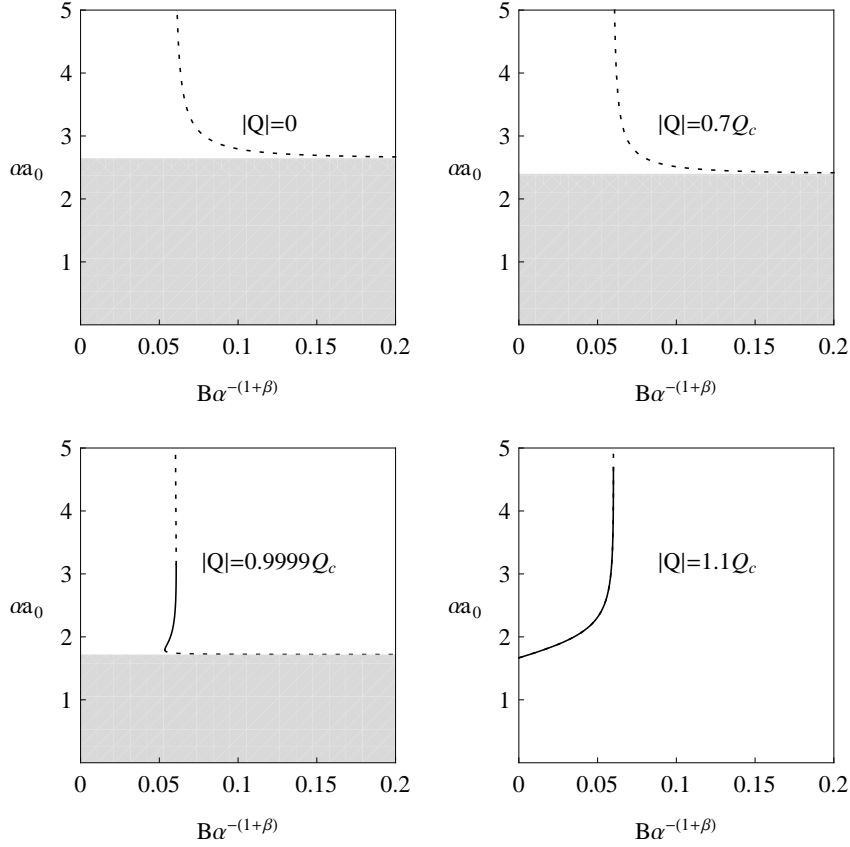}\caption{Charged black string
thin-shell wormholes for $\alpha=0.6,~A=0,~M=1$ and $\beta=0.2$.}
\end{figure}
When $\beta=0.2$, we have only unstable solution for
$|Q|=0,~0.7Q_c$, while for $|Q|=0.999Q_c$ there exist three
solutions two are unstable and one is stable. Finally, for
$|Q|>Q_c$, the horizon radius disappears and both stable and
unstable solutions exist similar to the above cases. For
$\beta=0.6$, the behavior of solutions presented in Figure
\textbf{5} is similar to the case $\beta=0.6$ for MGCG shown in
Figure \textbf{2} for $|Q|=0,~0.7Q_c, ~0.999Q_c$, while for
$|Q|>Q_c$, only stable solution exists in this case. When $\beta=1$,
we have only stable solutions for all values of $|Q|$ as shown in
Figure \textbf{6}.
\begin{figure}
\centering \epsfig{file=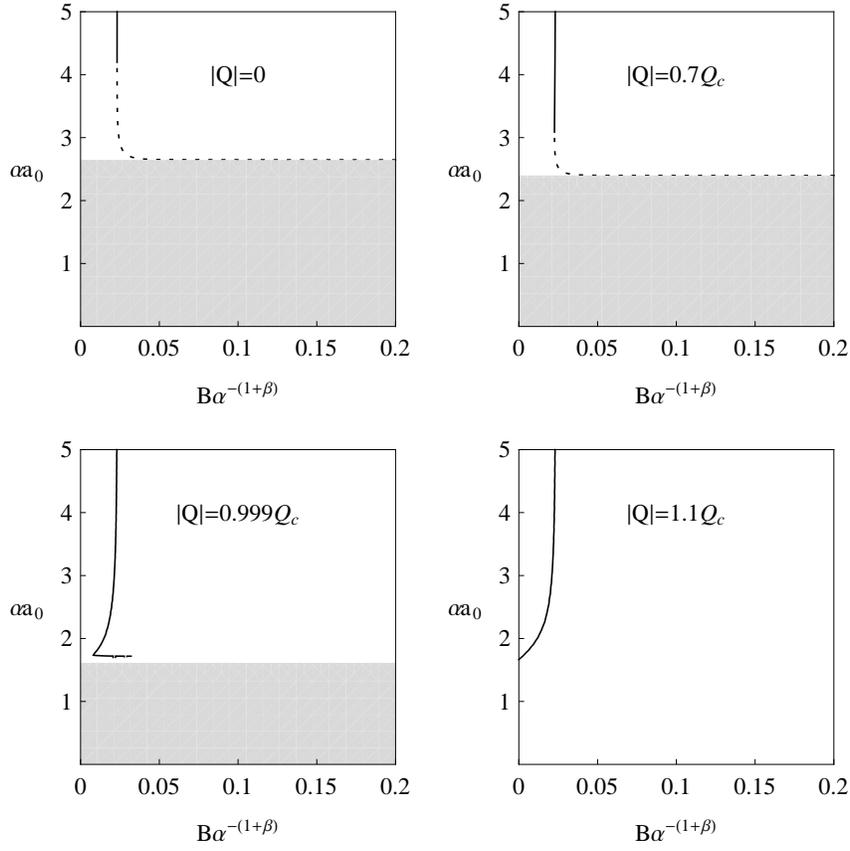}\caption{Charged black string
thin-shell wormholes for $\alpha=0.6,~A=0,~M=1$ and $\beta=0.6$.}
\end{figure}
\begin{figure}
\centering \epsfig{file=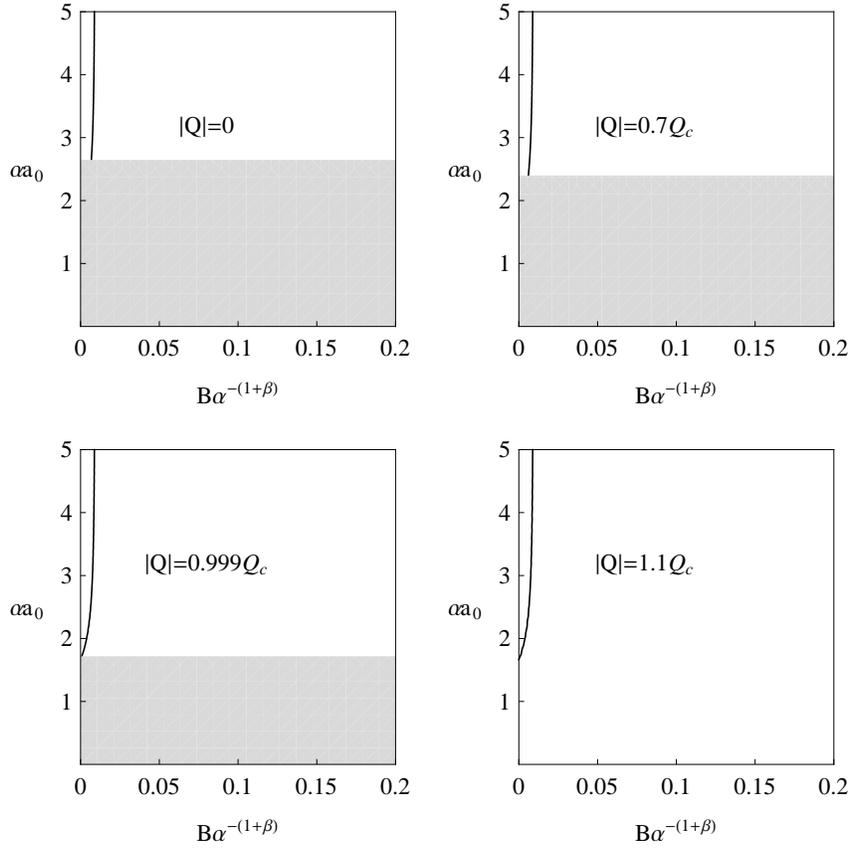}\caption{Charged black string
thin-shell wormholes for $\alpha=0.6,~A=0,~M=1$ and $\beta=1$.}
\end{figure}

\section{Summary}

The study of thin-shell wormholes has been the subject of interest
due to the presence of exotic matter, which violates the null energy
condition. The aim of this study is to construct cylindrical
thin-shell wormholes and investigate the stability of these
configurations. We have developed cylindrical thin-shell wormholes
by joining two identical copies of the cylindrical manifold using
cut and paste method. In order to explore the dynamics of thin-shell
wormhole, we have applied the Darmois-Israel junction conditions
along with MGCG for the description of exotic matter. We have
explored the stability of static solutions numerically satisfying
the condition $a_0>r_h$, under linear perturbations.

The stability of thin-shell wormhole solutions is examined for
different values of parameter $\beta=0.2,~0.6,~1$. It is found that
for $\beta=0.2,~0.6$, there always exist stable and unstable
solutions except for $\beta=0.6$ and $|Q|=0.999Q_c$, for which only
stable static solution exists. Moreover, in both cases the horizon
radius decreases with the increase of $|Q|$. When $\beta=1$, we have
stable configurations for small values of $B\alpha^{-(1+\beta)}$
with $|Q|<Q_c$ and approaches to the horizon of the manifold,
whereas for $|Q|>Q_c$, we have obtained stable as well as unstable
solutions like the other cases for $\beta=0.2,~0.6.$

Further, we examine the stability of solutions for the GCG. It is
found that for $\beta=0.2$, there exists one unstable solution with
$|Q|=0,0.7Q_c$ and two unstable and one stable solutions for
$|Q|=0.999Q_c$. Also, when $|Q|>Q_c$, there exist both stable and
unstable solutions. For $\beta=0.6$, the solutions are similar to
the case of MGCG with $|Q|=0,~0.7Q_c,~0.999Q_c$, while only stable
solution exists for $|Q|>Q_c$.

Recently, we have found stable configurations for the uncharged and
charged black string wormholes supported by Chaplygin gas \cite{a}.
We would like to mention here that the results of this work reduce
to \cite{a} for $\beta=1$ and $A=0$ as shown in Figure \textbf{6}.
It is worth mentioning here that the literature \cite{17}-\cite{20}
indicates only unstable solutions for cylindrical thin-shell
wormhole. The apparent discrepancy of our results with those
presented in the literature comes from the Einstein field equations
and the choice of the equation of state for the exotic matter. We
have concluded that there is a possibility of stable configuration
for the cylindrical thin-shell wormholes.\\

\vspace{0.5cm}

{\bf Acknowledgment}

\vspace{0.5cm}

We would like to thank the Higher Education Commission, Islamabad,
Pakistan, for its financial support through the {\it Indigenous
Ph.D. 5000 Fellowship Program Batch-VII}. One of us (MA) would like
to thank University of Education, Lahore for the study leave.

\end{document}